\documentclass{osa-article}
\journal{oe} 
\usepackage[T1]{fontenc}
\usepackage[utf8]{inputenc}
\usepackage{amsmath}
\usepackage{bm} 
\usepackage{soul}
\usepackage{graphicx}
\usepackage{bbm}
\usepackage{xspace}
\usepackage{bm}
\usepackage{booktabs}
\usepackage{babel}
\usepackage[separate-uncertainty=true]{siunitx}
\usepackage[obeyFinal,colorinlistoftodos,color=green]{todonotes} 
\usepackage{ifdraft}
\definecolor{purple}{rgb}{0.5,0,0.5}
\usepackage[normalem]{ulem}

\makeatletter

\usepackage[normalem]{ulem}

\begin{document}

\title{Calibration of Spin-Light Coupling by Coherently Induced Faraday Rotation}
\author{Rodrigo A. Thomas,\authormark{1,*} Christoffer Østfeldt,\authormark{1} Christian Bærentsen,\authormark{1} Michał Parniak,\authormark{1,2} Eugene S. Polzik\authormark{1}}

\address{\authormark{1}Niels Bohr Institute, University of Copenhagen, Copenhagen, Denmark \\
\authormark{2}Centre for Quantum Optical Technologies, Centre of New Technologies, University of Warsaw, Warsaw, Poland}

\email{\authormark{*}rathomas@nbi.ku.dk} 

\begin{abstract}
Calibrating the strength of the light-matter interaction is an important experimental task in quantum information and quantum state engineering protocols. The strength of the off-resonant light-matter interaction in multi-atom spin oscillators can be characterized by the coupling rate $\Gamma_\text{S}$.  Here we utilize the Coherently Induced Faraday Rotation (CIFAR) signal for determining the coupling rate. The method is suited for both continuous and pulsed readout of the spin oscillator, relying only on applying a known polarization modulation to the probe laser beam and detecting a known optical polarization component. Importantly, the method does not require changes to the optical and magnetic fields performing the state preparation and probing. The CIFAR signal is also independent of the probe beam photo-detection quantum efficiency, and allows direct extraction of other parameters of the interaction, such as the tensor coupling $\zeta_\text{S}$, and the damping rate $\gamma_\text{S}$. We verify this method in the continuous wave regime, probing a strongly coupled spin oscillator prepared in a warm cesium atomic vapour.
\end{abstract}

\ifoptionfinal{}{
\section{Introduction}
}{}

The off-resonant interface of light with atomic ensembles has been widely explored in the last decades \cite{HapperOpticalPumping,budker2007optical,EchanizPRA2008,RMPKlemensAndersEugene} in ultra-cold, cold and warm alkali implementations. The spin degree of freedom in the atomic ground state coherences and its coupling to light has been used in protocols ranging from fundamental \cite{julsgaard2001experimental,julsgaard2004experimental,VasilakisPRL2011,krauter2013deterministic} to technological \cite{AcostaNMOR2006,Wasilewski2010,jensen2016non} applications. Furthermore, the collective spin excitations of the highly polarized atomic ensembles in a static magnetic field can be well approximated by harmonic oscillator-like degrees of freedom -- a spin oscillator \cite{polzikadp2015}. The oscillator mapping
is essential for the interface with nano-mechanical oscillators via back-action evasion \cite{Moller2017} and entangling \cite{Thomas2020} measurements, which promises sensitivity improvements in future gravitational wave detectors \cite{ZeuthenGWD2019} and optical quantum control of the hybrid spin-mechanical system \cite{karg2020light}.

In the interface between atoms and light, characterizing the strength with which the systems couple is paramount for understanding its dynamics. According to the principles of quantum mechanics, the statistical nature of the quantum measurement process leads to fundamental limits in estimation of system's state at a given instant of time \cite{braginsky1992quantum}. Optimizing the role of the measurement performed by the optical probe on the spin oscillator is key.

The optical readout of a highly polarized atomic ensemble prepared in a spin oscillator state contains contributions arising from \cite{AuzinshPRL2004}: optical shot noise, inherent to the quantum nature of light in the process of photo-detection; coherent spin state or ground state noise, from the zero-point energy required to satisfy the Heisenberg uncertainty principle; thermal noise, originating from the extra spin fluctuations of the oscillator having non-zero mean occupancy $n_\text{S}$ due to imperfect optical pumping; and lastly, quantum back-action noise, originating from the perturbations of the optical readout in the oscillator's dynamics. As we obtain information about the oscillator by performing measurements on the light that has interacted with the system, it is of key importance to faithfully characterize the weight of each of these contributions. 

The standard quantum limit (SQL) for a measurement of mechanical displacements, for example, sets the sensitivity to external fields in conventional interferometric measurements \cite{braginsky1967sql}. At the SQL, the detection shot noise and measurement back-action contribute equally to the measurement imprecision. A highly efficient mapping of the oscillator state to light requires the quantum back-action to dominate over the coupling to the thermal environment \cite{RMPKlemensAndersEugene}. Common to these protocols is the importance of the interaction strength parameter between light and the oscillator, here defined as the readout rate $\Gamma_\text{S}$ (also commonly known as the measurement rate in the optomechanics community \cite{rossi2018measurement}). Another figure of merit that quantifies the efficiency of the coupling is the quantum cooperativity $C_\text{q}$, here defined as $C_\text{q}=\tfrac{\Gamma_\text{S}}{2\gamma_\text{S}(n_\text{S}+1/2)}$. Working in the regime $C_\text{q}\gg 1$, in which the coupling is strong, indicates that the measurement significantly influences the oscillator dynamics, allowing for its control and manipulation. 

In this paper, we show how the parameter $\Gamma_\text{S}$ may be extracted from a measurement based on the interference of the induced Faraday rotation, i.e., the oscillator response to a classical optical polarization modulation, with the modulation itself. We call it Coherently Induced FAraday Rotation, or CIFAR, signal. The method further allows extraction of the effective linewidth, $\gamma_\text{S}$, and the tensorial part of the interaction, $\zeta_\text{S}$, describing the deviation from the idealized quantum non-demolition (QND) interaction. Crucially this method relies on the same alignment of magnetic and optical fields as well as optical pumping of the atomic ensemble, as used for experiments such as \cite{Moller2017, Thomas2020}. The CIFAR method is applicable in all experimental implementations of spin oscillator, from ultra-cold to warm vapors, in ensembles with total angular momentum (per atom) equal or larger than $\tfrac{1}{2}$.  The coherent drive also allows for probing the atomic motion  through the laser beam \cite{Borregaard2016} and characterizing the coupling to fast decaying spin modes \cite{Shaham2020}, which give rise to a broadband spin response. We verify the CIFAR in the continuous regime, probing a spin oscillator in the strong coupling regime, prepared in a warm cesium atomic vapour. Lastly, we also present the limits of our linearized oscillator description by driving the system with large classical polarization modulation.

The technique here described is especially suited for continuous wave measurement of (single) spin oscillators, which should be contrasted with the mean value transfer method \cite{Wasilewski2010} and thermal noise scaling \cite{hannathesis}, which rely on time consuming measurements and more dramatic changes of the experimental setup. We nonetheless highlight that this technique can also be employed in the canonical state-preparation-probing experimental cycle. Furthermore, the signal depends only on the interference between the drive and response, it is independent of the overall detection quantum efficiency and thermal noise calibrations, as the calibration used in \cite{Moller2017}. The CIFAR method is similar to the Optomechanically Induced Transparency (OMIT)-response measurement technique \cite{OMITScience2010}, used to characterize optomechanical coupling parameters \cite{NielsenpPNAS2017}.

\ifoptionfinal{}{
\section{Theory}
}{}
\label{sec:model}

The interaction between an atomic ensemble and light has been widely studied in the context of optical pumping \cite{HapperOpticalPumping} and quantum information applications \cite{RMPKlemensAndersEugene}. For a far detuned monochromatic optical field with intensity much below saturation, the effective atom-light interaction is given by the coupling of the electronic-ground-state magnetic sublevels to the light polarization. The interaction can be seen as a mutual rotation of the light and spin variables, $ \{\hat{S}_x,\hat{S}_y,\hat{S}_z,\hat{S}_0\}$ and $ \{\hat{J}_x,\hat{J}_y,\hat{J}_z,\hat{J}_0\}$ respectively, in the form of polarization-dependent ac Stark shifts of the ground state levels and spin-state-dependent index of refraction, according to the atomic polarizability tensor~\cite{julsgaardthesis,deutsch2010quantum}. Typical ensembles consists of $N\approx10^8-10^{12}$ atoms, with the collective macroscopic spin being represented as $\hat{J}_{x,y,z} = \sum_{i=1}^N\hat{F}^{(i)}_{x,y,z}$, where $\hat{F}_{x,y,z}^{(i)}$ are the Cartesian decomposition of the total angular momentum operator of a single atom.

The light-matter interaction, along with the contribution from an external bias magnetic field applied in the $x$-direction, gives the spin Hamiltonian 
\begin{align}\label{eq:Hs0}
    \hat{H}_\text{S}/\hbar &= \pm\omega_\text{S} \hat{J}_x + g_\textrm{S} \bigg[ a_0 \hat{S}_0 \hat{J}_0 + a_1 \hat{S}_z\hat{J}_z +2a_2 \left[ \hat{S}_0\hat{J}_z^2 - \hat{S}_{x}(\hat{J}_{x}^2-\hat{J}_{y}^2) - \hat{S}_{y}(\hat{J}_x\hat{J}_y + \hat{J}_y \hat{J}_x)\right]\bigg],
\end{align}
where the first term refers to the magnetic coupling induced bias magnetic field, shifting the ground state magnetic sublevels by $\pm\omega_\text{S}$, with sign depending on the direction of the magnetic field with respect to the mean spin. The coefficients $a_0$, $a_1$, and $a_2$ as the relative weights of the scalar, vector and tensor contributions of the polarizability tensor \cite{deutsch2010quantum}, and $g_\textrm{S}$ is the single-photon coupling rate. The relative weights of the contributions depend on the level structure of the atom and can be controlled by the laser detuning from the atomic resonance. The vector and tensor contributions are related to circular and linear birrefringence of the atomic medium, respectively. The scalar component leads to a polarization independent phase shift. 

We now focus in the specific case of cesium \cite{julsgaardthesis}. For a laser beam detuned by $\Delta$ from the $F=4\rightarrow F'=5$ transition in the D$_2$ cesium line interacting with atoms in the $F=4$ ground state manifold, the a$_i$ parameters are given by
\begin{equation}
    \begin{split}
        a_{0}&= \frac{1}{4}\left(\frac{1}{1+\Delta_{35} / \Delta}+\frac{7}{1+\Delta_{45} / \Delta}+8\right)  \\
        a_{1}&= \frac{1}{120}\left(-\frac{35}{1+\Delta_{35} / \Delta}-\frac{21}{1+\Delta_{45} / \Delta}+176\right) \\
        a_{2}&= \frac{1}{240}\left(\frac{5}{1+\Delta_{35} / \Delta}-\frac{21}{1+\Delta_{45} / \Delta}+16\right),
    \end{split}
\end{equation}
with $\Delta_{35}/2\pi=\SI{452}{\mega\hertz}$ and $\Delta_{45}/2\pi=\SI{251}{\mega\hertz}$ as the excited state hyperfine splittings between $F'=3$ and $F'=5$, and $F'=4$ and $F'=5$ \cite{steck2003cesium}, respectively. A detuning $\Delta>0$ ($\Delta<0$) corresponds to the case with laser frequency above (below) the $F=4\rightarrow F'=5$ transition.  In our experiments, the probe laser is detuned by $\Delta/2\pi=\SI{3}{\giga\hertz}$, where relative weights are $a_0 \sim 3.83$, $a_1 \sim 1.05$, and $a_2\sim 0.004$.

In practice, the ensemble is not perfectly polarized due to limited optical pumping efficiency and decay due to, e.g.,~wall collisions, natural lifetime and optical de-pumping. Since the values for $a_i$ above are calculated for perfect spin polarization, the effective values observed experimentally differ somewhat from those stated here. Further, the imperfect spin polarization gives rise to a thermal, stochastic distribution of the spins in the different magnetic sublevels, which shows as thermal noise in the detection. 

The interaction between light and the atomic ensemble in equation~\eqref{eq:Hs0} can be simplified and linearized in the case of large ground state spin polarization and a strong polarized classical laser field. These approximations constitute the mapping of the spin system to an oscillator system. As we will describe in the next section, the ensemble is optically pumped such that $ J_x = \langle \hat{J}_x \rangle $ and $\langle \hat{J}_y \rangle,\langle \hat{J}_z \rangle \ll J_x $ at any instant of time. 

The probe is a strong classical field linearly polarized at an angle $\alpha$ to the magnetic field, which is also the quantization axis. The angle $\alpha$ controls the relative contributions of the vector and tensor effects described by the Hamiltonian \eqref{eq:Hs}. For simplicity, we change basis of the polarization variables such that the component parallel to the quantization axis, here $\hat{S}_\parallel$, describes the strong field as a classical variable with mean photon flux $\langle \hat{S}_\parallel \rangle = \langle \hat{S}_0 \rangle = S_\parallel$, leaving ${\hat{S}_\perp, \hat{S}_z}$ as zero-mean quantum variables. Mathematically, we rotate the polarization variables around the $\hat{S}_z$ components as $\{\hat{S}_\parallel=\hat{S}_x\cos 2\alpha-\hat{S}_y\sin 2\alpha,\hat{S}_\perp = \hat{S}_x\sin 2\alpha+\hat{S}_y\cos 2\alpha,\hat{S}_z,\hat{S}_0\}$. 

For a highly polarized ensemble in the $F=4$ hyperfine manifold, the Hamiltonian in equation \eqref{eq:Hs0} can be simplified \cite{WasilewskiOE2009}. In the limit of high steady state spin polarization, where only the two extreme magnetic sublevels, i.e.~either $m_F=+4,+3$ or  $m_F=-4,-3$, are populated, we perform the Holstein-Primakoff approximation \cite{holsteinprimakoff} and map the spin variables to effective position and momentum variables
\begin{align}\label{eq:Hs}
    \hat{H}_\text{S}/\hbar &=\mp\dfrac{ \omega_\text{S} }{2}(\hat{X}_\text{S}^2+\hat{P}_\text{S}^2) - 2\sqrt{\Gamma_\text{S}}\left(\hat{X}_\text{S} \hat{X}_\text{L} + \zeta_{\text{S}} \hat{P}_\text{S} \hat{P}_\text{L} \right).
\end{align}
The canonical variables for light and spins are $\{\hat{X}_\text{L}=\hat{S}_\text{z}/\sqrt{S_\parallel},\hat{P}_\text{L}=-\hat{S}_\perp/\sqrt{S_\parallel}\}$ and $\{\hat{X}_\text{S}=\hat{J}_\text{z}/\sqrt{|J_x|},\hat{P}_\text{S}=-\text{sgn}(J_x)\hat{J}_y/\sqrt{|J_x|}\}$, respectively, satisfying $[\hat{X}_{\text{L}}(t),\hat{P}_{\text{L}}(t')]=(i/2)\delta(t-t')$ and $[\hat X_\text{S},\hat P_\text{S}]=i$. The quantity $\text{sgn}(J_x)$ refers to the sign of the mean spin, being positive (negative) for the negative (positive) mass oscillator cases.Notice that the sign of $\omega_\text{S}$ carries information about the mutual orientation of $J_x$ and the external bias magnetic field. In the harmonic oscillator language, the mutual orientation defines the effective mass of the spin oscillator, with $-\omega_\text{S}$ ($+\omega_\text{S}$) referring to the negative (positive) mass.  In the derivation of equation~\eqref{eq:Hs}, we have omitted constant energy terms, as they do not affect the dynamics of the variables of interest. 

The parameters
\begin{align}
    \Gamma_\text{S} &= g_\textrm{S}^2a_1^2 S_\parallel J_x \label{eq:readoutrate}\\
    \zeta_{\text{S}} &= -14  \frac{a_2}{a_1} \cos2\alpha \label{eq:tornado},
\end{align}
are the spin oscillator readout rate and the tensor coupling parameter, respectively. If $\zeta_\text{S} = 0$ the light-spins interaction is of the Quantum Non-Demolition (QND) type. In our experimental regime, as $\zeta_\text{S} \neq 0$, the spin-light interactions deviates from the QND interaction, allowing for dynamical cooling/heating of the spin oscillator and changing the total decay rate and effective bath occupation in similar fashion to the effects of light interaction with a mechanical oscillator in the field of optomechanics \cite{cavityoptomechRMP}. Experimentally, due to imperfect spin polarization, we will have $\zeta_\text{S}$ smaller than the value predicted from equation \eqref{eq:tornado}. As shown in Section \ref{sec:experiments}, our full model with $\zeta_\text{S}$ as free parameter describes the measured response sufficiently well.

The spin system evolves coherently due to the Hamiltonian given in equation \eqref{eq:Hs}, and incoherently due to spin decay and coupling to an external effective spin bath \cite{vasilyev2012quantum}. Furthermore, atomic motion leads to a time-dependent light-spin coupling. There is, in principle, an infinite set of collective spin modes that evolve in time accordingly to the ensemble geometry, collisions, dephasing, and diffusion \cite{Shaham2020}. Here, we focus on the so called \textit{flat} spin mode corresponding to the total spin $\hat{J}_k=\sum_{i=1}^N \hat{F}_k^{(i)}$, 
the mode which is the most resilient to motional dephasing as it is fully symmetric with respect to shuffling atomic positions. In the linearized language introduced above, we assign effective position and momentum variables, $  \hat X_\text{S}, \hat P_\text{S} $ to this mode. Later on, we will also introduce effective variables to describe the dynamics of the fast decaying spin modes, here denominated as the \textit{broadband response}, and also introduce a qualitative model that describes its response to light.

The dynamics of the spin variables due to the Hamiltonian presented in equation \eqref{eq:Hs}, for the case with effective positive mass, is
\begin{align}\label{eq:evomatrix}
    \dfrac{\mathrm{d}}{\mathrm{d}t}\begin{pmatrix}
    \hat{X}_{\text{S}}\\
    \hat{P}_{\text{S}}
    \end{pmatrix} = \begin{pmatrix}
    -\gamma_{\text{S}}/2 & \omega_{\text{S}}\\
    -\omega_{\text{S}} & -\gamma_{\text{S}}/2
    \end{pmatrix} \begin{pmatrix}
    \hat{X}_{\text{S}}\\
    \hat{P}_{\text{S}}
    \end{pmatrix} +2\sqrt{\Gamma_\text{S}}  \begin{pmatrix}
    0 & -\zeta_{\text{S}}\\
    1 & 0
    \end{pmatrix} \begin{pmatrix}
    \hat{X}_{\text{L}}^\textrm{in}\\
    \hat{P}_{\text{L}}^\textrm{in}
    \end{pmatrix},
\end{align}
with $\gamma_{\text{S}}/2=\gamma_{\text{S0}}/2+\zeta_{\text{S}}\Gamma_{\text{S}}$ as the dynamical damping rate, including tensor effects \cite{kasperthesis}. Here, $\gamma_{\text{S0}}$ includes the natural (in the dark) decay rate, and laser induced contributions (from the pumping and probing lasers). For notation purposes, we write the light and spin variables in the matrix form as
\begin{align}
     {\mathbf{X}}_{\text{L}}^\textrm{in(out)}=\begin{pmatrix}
    {X}_{\text{L}}^\textrm{in(out)}\\
    {P}_{\text{L}}^\textrm{in(out)}
    \end{pmatrix}, \qquad 
     {\mathbf{X}}_{\text{S}}=\begin{pmatrix}
    {X}_{\text{S}}\\
    {P}_{\text{S}}
    \end{pmatrix},
\end{align}
where the superscripts \textit{in} (\textit{out}) denote the optical mode before (after) the interaction with the spin oscillator, to be presented below. In the CIFAR experiments, as we will discuss in Section \ref{sec:experiments}, the oscillator is coherently excited, e.g. with a drive $X_\textrm{L}^{\textrm{in}} \propto  \sin\omega_\text{RF}t $. Since the system is driven by a classical driving field, the system response can also be considered as a classically, and we drop the operator description from here on.

Given the linear system of equations \eqref{eq:evomatrix} and a sinusoidal drive input ${\mathbf{X}}_{\text{L}}^\textrm{in}$, a solution in the complex plane can be found using the ansatz $X_\text{S}(t)=X_\text{S}(\omega_\text{RF})e^{-i\omega_\text{RF}t}, P_\text{S}(t)=P_\text{S}(\omega_\text{RF})e^{-i\omega_\text{RF}t}$, where $X_\text{S}(\omega_\text{RF})$ and $X_\text{S}(\omega_\text{RF})$ are complex numbers. 
We write this as
\begin{align}\label{eq:Xs}
     {\mathbf{X}}_\text{S}&=2\sqrt{\Gamma_\text{S}}\mathbf{L}\mathbf{Z} {\mathbf{X}}_\text{L}^\text{in},
\end{align}
where the matrices $\mathbf{L}$ and $\mathbf{Z}$ parametrize the interaction dynamics as
\begin{align} 
    \mathbf{Z}&=\begin{pmatrix}
    0 & -\zeta_{\text{S}}\\
    1 & 0
    \end{pmatrix}\label{eq:Zdef} \\
    \mathbf{L}&=\begin{pmatrix}
    \gamma_{\text{S}}/2-i\omega_\text{RF} & -\omega_{\text{S}}\\
    \omega_{\text{S}} & \gamma_{\text{S}}/2-i\omega_\text{RF}
    \end{pmatrix}^{-1}
    =
    \chi_\text{S}(\omega_\text{RF})
    \begin{pmatrix}
    \gamma_\text{S}/2-i\omega_\text{RF} & \omega_\text{S} \\
    -\omega_\text{S} & \gamma_\text{S}/2-i\omega_\text{RF}
    \end{pmatrix},
    \label{eq:Ldef}
\end{align}
with $ \chi_\text{S}(\omega_\text{RF})=\left(\omega _\text{S}^2+\left(\frac{\gamma _\text{S}}{2}-i \omega_\text{RF} \right)^2\right)^{-1} $ as the spin susceptibility. All spin and light variables are understood to be functions of drive frequency, $\omega_\text{RF}$, the notation of which we suppress from now on.

The output light field, after the interaction with the spin oscillator given in equation \eqref{eq:Xs}, is 
\begin{align}\label{eq:XLoutcifar}
{\mathbf{X}}_\text{L}^\text{out}&={\mathbf{X}}_\text{L}^\text{in} + \sqrt{\Gamma_{\text{S}}}\mathbf{Z}{\mathbf{X}}_\text{S}=(\mathbf{1}_2+2\Gamma_\text{S}\mathbf{Z} \mathbf{L}\mathbf{Z}){\mathbf{X}}_\text{L}^\text{in},
\end{align}
where $\mathbf{1}_2$ is the $2\times 2$ identity matrix. The equation above shows that the output light field will have two contributions: one directly from the input field and another from the response of the spin oscillator to the input. 
Having the light field as a common source, these two contributions can interfere. 

By inserting \eqref{eq:Zdef} and \eqref{eq:Ldef} into \eqref{eq:XLoutcifar}, we get the expressions for output optical quadratures after the interaction with the spin ensemble
\begin{align}\label{eq:inout}
    \begin{pmatrix}
    X^{\text{out}}_\text{L} \\
    P^{\text{out}}_\text{L}
    \end{pmatrix} &= \begin{pmatrix}
     1-2 \Gamma _\text{S} \zeta_\text{S} \left(\tfrac{\gamma _\text{S}}{2}-i \omega_\text{RF} \right)\chi_\text{S}(\omega_\text{RF}) & -2 \Gamma _\text{S} \zeta_\text{S}^2 \omega _\text{S} \chi_\text{S}(\omega_\text{RF})\\
     2 \Gamma _\text{S} \omega _\text{S} \chi_\text{S}(\omega_\text{RF}) & 1-2 \Gamma _\text{S} \zeta_\text{S} \left(\tfrac{\gamma _\text{S}}{2}-i \omega_\text{RF} \right)\chi_\text{S}(\omega_\text{RF})
    \end{pmatrix}
    \begin{pmatrix}
    X^{\text{in}}_\text{L} \\
    P^{\text{in}}_\text{L}
    \end{pmatrix}.
\end{align}

In general, we are able to select arbitrary input $ {\mathbf{X}}_\text{L}^\text{in} $ and the detection $ {\mathbf{X}}_\text{L}^\text{out} $ components by controlling the ellipticity of the polarization state by the phases $\theta$ and $\phi$, respectively. The input light state, without loss of generality, is assumed to be generated from a pure phase modulation $G=|G|e^{i\varphi}$ that, when referenced to a local oscillator (LO) in a Mach-Zehnder interferometer, as we have in Figure~\ref{fig:setup}, can be arbitrarily decomposed into polarization variables and effective input amplitude and phase quadratures. Here, by convention, we have chosen ${X}^{\text{in}}_\text{L}=G,{P}^{\text{in}}_\text{L}=0$ for $\theta=0$. Path length difference control in the Mach-Zehnder interferometer allows for mixing the drive components via a basis rotation, and polarization homodyning allows for selecting the detection quadrature, as
\begin{align}\label{eq:rotations}
\begin{pmatrix}
{X}^{\text{in}}_\text{L} \\
{P}^{\text{in}}_\text{L}
\end{pmatrix} &= \begin{pmatrix}
\cos\theta & -\sin\theta \\
\sin\theta & \cos\theta
\end{pmatrix} \begin{pmatrix}
G \\
0
\end{pmatrix} = \begin{pmatrix}
\cos\theta  \\
\sin\theta
\end{pmatrix} G, \qquad
\begin{pmatrix}
{X}^{\text{det}}_\text{L} \\
{P}^{\text{det}}_\text{L}
\end{pmatrix} = \begin{pmatrix}
\cos\phi & -\sin\phi \\
\sin\phi & \cos\phi
\end{pmatrix} \begin{pmatrix}
{X}^{\text{out}}_\text{L} \\
{P}^{\text{out}}_\text{L}
\end{pmatrix}.
\end{align}
By inserting the equations \eqref{eq:rotations} in equation \eqref{eq:inout}, we get to the final form of the Coherently Induced Faraday Rotation (CIFAR) signal. We typically define the measured quadrature as $ P^{\text{det}}_\text{L} $, such that the absolute squared of the detected spin response to an arbitrary input optical modulation is
\begin{align}
     |\text{CIFAR}|^2 \equiv \left|{P}^{\text{det}}_\text{L}\right|^2 &= \left|{P}^{\text{out}}_\text{L}\cos\phi+{X}^{\text{out}}_\text{L}\sin\phi \right|^2 \nonumber \\
    &= \left|\left(1-2 \Gamma _\text{S} \zeta _\text{S} \left(\tfrac{\gamma _\text{S}}{2}-i \omega_\text{RF} \right)\chi_\text{S}(\omega_\text{RF})\right)\sin(\theta+\phi) \right. \nonumber \\
    & \left. + \medspace \Gamma _\text{S} \omega _\text{S} \chi_\text{S}(\omega_\text{RF}) \left[ (1-\zeta_\text{S}^2) \cos(\theta-\phi) + (1+\zeta_\text{S}^2) \cos(\theta+\phi) \right] \right| ^2 |G|^2 \label{eq:CIFARfinal}.
\end{align}
This equation is the main result of this section, being applicable to the description of the flat spin mode response to light.

In a broader view, it becomes necessary to consider other spin modes which in contrast to the total spin will have some spatial dependencies. We consider collective operators corresponding to the transverse spin components of mode $n$ given by $\hat{J}_{z,y}^n=\sqrt{V} \sum_{i=1}^N  u_n(\mathbf{x}_i) \hat{F}_{z,y}$ where $u_n(\mathbf{x})$ represents the spatial shape of the spin mode, $V$ is the volume of cell (for the purpose of proper normalization) and $\mathbf{x}_i$ is the position of $i$-th atom. The coherent evolution of each mode (collective operator) is then governed by equation \eqref{eq:Hs} with the readout rate $\Gamma_\mathrm{S}^n$ now taking into account the overlap between the spin mode and the Gaussian light mode $I^\mathrm{G}_n$, such that $\Gamma_\mathrm{S}^n \sim |I^\mathrm{G}_n|^2$ \cite{Shaham2020}. At the same time, the incoherent part will depend on motion and wall collisions. It has been shown that even in the paraffin- or alkene-coated cells \cite{Sekiguchi2016,Hatakeyama2019,Tang2020,Atutov2017} or cells with dilute buffer gas \cite{Parniak2014} the atomic motion can be effectively described by the diffusion equation with adequate wall boundary condition. The cells with coated walls will feature a slow decay for the flat mode ($u_0(\mathbf{x}_i)=1/\sqrt{V}$) which depends on intrinsic dynamics and wall decay, and much faster decay for all other modes, which given by $\gamma_\mathrm{S}^n = D k_n^2$, where $D$ is the effective diffusion constant and $k_n$ is the characteristic wavenumber of $n$-th mode \cite{Shaham2020}. 

For the case of quantum noise, it becomes necessary to consider both thermal contributions of each mode (added incoherently), and the coherent interaction of each mode the the Gaussian beam. These broadband spin noises affect the atomic ensembles serving as magnetometers \cite{Lucivero2017} or quantum memories \cite{Shaham2020}. In our case we can consider only the coherent interaction and thus each spatial spin mode responds to the same light modulation. Therefore, we may simply modify equation \eqref{eq:XLoutcifar} to the multimode case:
\begin{align}\label{eq:XLoutwithbb}
{\mathbf{X}}_\text{L}^\text{out}&=\left(\mathbf{1}_2+\sum_n 2\Gamma_\text{S}^n\mathbf{Z} \mathbf{L}_n\mathbf{Z}\right){\mathbf{X}}_\text{L}^\text{in},
\end{align}
with $\mathbf{L}_n$ contains the susceptibility with the respective decay rate $\gamma_\mathrm{S}^n$. In our case we shall work in a two-mode approximation for which the zeroth mode is the flat, long-lived mode, and the other mode has $\gamma_\mathrm{S}/2\pi \sim \SI{1}{\mega\hertz}$. This is a justified approach as all broad modes contribute a similar flat background around the resonance which we primarily study here. Following this approach we obtain a two-component CIFAR signal in which the narrow (with linewidth $\gamma_\mathrm{S}$ and readout rate $\Gamma_\mathrm{S}$) and broad parts (with linewidth $\gamma_\mathrm{S,BB}$ and readout rate $\Gamma_\mathrm{S,BB}$) of the response may interfere according to their phase relation. In the following sections we will nevertheless give simplified formulas for the single mode case to facilitate understanding, and use the two-component model for fitting.

Given the input and detection angles, as well as the spin oscillator parameters and coupling to light, the CIFAR signal exhibits a characteristic frequency response. For developing intuition about the response, let us focus initially in the special case of $ \theta=\SI{45}{\degree} $ and $ \phi=\SI{0}{\degree} $, corresponding to detecting the phase quadrature of light $ {P}^{\text{out}}_\text{L} $ and driving with an equal superposition of amplitude $ {X}^{\text{in}}_\text{L} $ and phase modulation $ {P}^{\text{in}}_\text{L} $. For the choice of phases described above, the detected signal goes as 
\begin{align}\label{eq:cifarpout}
|\text{CIFAR}(\theta=\SI{45}{\degree}, \phi=\SI{0}{\degree}) |^2 = \left|1- 2\Gamma_\text{S} (-\omega_\text{S} +\zeta_\text{S}  (\gamma_\text{S}/2- i \omega_\text{RF} ))\chi_\text{S}(\omega_\text{RF})\right|^2|G|^2.
\end{align}
Notably, the drive, represented by the constant term, and the spin response, proportional to the susceptibility $ \chi_\text{S}(\omega_\text{RF}) $, are added coherently and interfere. In particular, the readout rate $\Gamma_\text{S}$ plays an important role in the interference pattern, modulating its strength. In the high-Q limit ($ \gamma_\text{S} \ll \omega_\text{S}$) and around resonance ($\omega_\text{RF} \sim \omega_\text{S}$), the spin susceptibility is $ \chi_{\text{S}} \sim -\chi_{\text{S}0}/\omega_\text{S} $, for $ \chi_{\text{S}0} =\tfrac{1}{2}((\Delta_\text{RF}+i \gamma_\textrm{S}/2))^{-1} $, with $ \Delta_\text{RF} = \omega_\text{RF}-\omega_\text{S} $ as the detuning between the spin resonance and the input modulation tone. In this limit, the equation \eqref{eq:cifarpout} becomes
\begin{align}\label{eq:cifarexp}
|\text{CIFAR}|^2 &\sim |1- 2\Gamma_\text{S}(1+i\zeta_\text{S})\chi_{S0}(\omega_\text{RF})|^2 = 1 + \dfrac{\Gamma_\text{S}^2(1+\zeta_\text{S}^2)-2\Gamma_\text{S}(\Delta_\text{RF}+\zeta_\text{S}\gamma_\text{S})}{\Delta_\text{RF}^2+ (\gamma_\text{S}/2)^2}.
\end{align}
For exemplifying the procedure to extract the readout rate parameter $\Gamma_\text{S}$, let us consider two specific cases. First, we analyze the case of $\zeta_\text{S} =0 $, that is, the light-matter interaction is of the QND type. Here,equation \eqref{eq:cifarexp} reduces to
\begin{align}\label{eq:cifarpoutQND}
|\text{CIFAR}_0|^2 &= 1 + \dfrac{\Gamma_\text{S}^2-2\Gamma_\text{S}\Delta_\text{RF}}{\Delta_\text{RF}^2+ (\gamma_\text{S0}/2)^2}.
\end{align}
The CIFAR$_0$ signal is a combination of a constant, a Lorentzian, and a dispersive term, representing the drive, the spin response and the interference between the drive and response, respectively. Importantly, the minimum and maximum of the signal are separated by $\sim  \sqrt{\Gamma_\text{S}^2+\gamma_\text{S} ^2} \sim \Gamma_\text{S}$
in the limit of high coupling, $\Gamma_\text{S}\gg \gamma_\text{S}$. The readout rate can thus be extracted just by noting this frequency difference, directly from the sweep figure. 

For the second specific case, when $\zeta_\text{S} \neq 0 $, equation \eqref{eq:cifarexp} leads to a correction of the separation, as the maximum and minimum are separated by $\sim  \sqrt{(1+\zeta_\text{S}^2)(\Gamma_\text{S}^2(1+\zeta_\text{S}^2)+\gamma_\text{S}^2-2\Gamma_\text{S}\gamma_\text{S}\zeta_\text{S})}$. In the high-coupling limit, $\Gamma_\text{S}\gg\gamma_\text{S}$, this simplifies to $\sim \Gamma_\text{S}(1+\zeta_\text{S}^2)$.

Having derived the needed expressions, we now turn to an experimental investigation of the CIFAR signal under different situations.

\ifoptionfinal{}{
\section{Experimental implementation}
}{}
\label{sec:experiments}

\begin{figure*}
    \centering
    \includegraphics{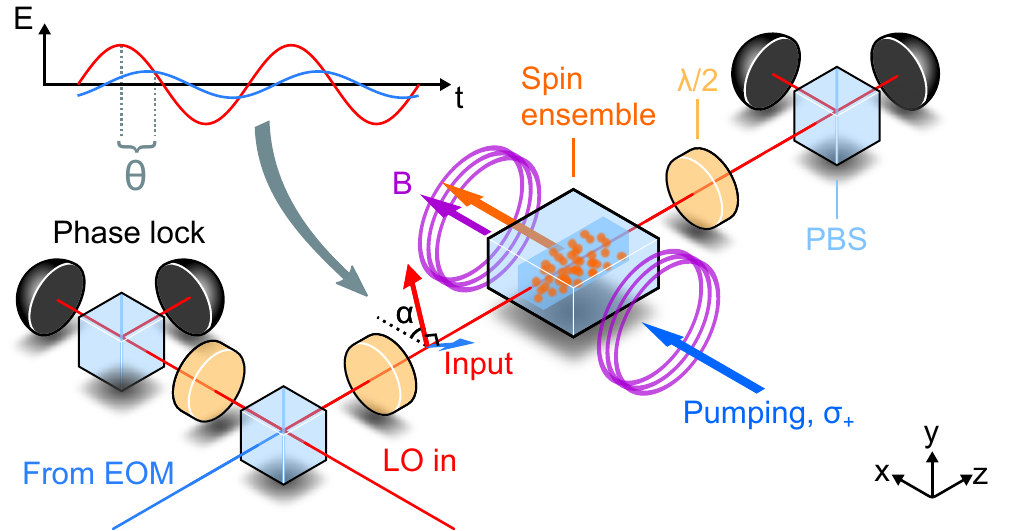}
    \caption{\textbf{Experimental setup.} A strong linearly polarized LO (red) is mode matched to a weaker, phase modulated beam from an EOM (blue) on a PBS. Part of the light is sent to a polarization sensitive detection setup, which is used to stabilize the relative phase between the two beams. The atomic input is sent through an optically polarized room temperature alkali spin ensemble (orange circles), situated in a homogeneous magnetic field $B$ (purple). The collective total spin vector (orange) modulates the input light polarization due to the Faraday interaction, while the light also drives the spin. The field at the output of the ensemble is detected in a polarization self-homodyning setup. EOM: Electro-optic modulator. LO: Local Oscillator. PBS: Polarizing beamsplitter. $\lambda/2$: Half wave plate.}
    \label{fig:setup}
\end{figure*}

The experimental setup is depicted in Figure~\ref{fig:setup}. We start by describing the atomic spin ensemble. The atomic spin ensemble is a warm gas of cesium atoms, confined to a spin anti-relaxation-coated microcell \cite{Balabas2010} with a $\SI{300}{\micro\meter}\times\SI{300}{\micro\meter}$ cross-section and $\SI{10}{\milli\meter}$ in length. The system is probed with a Gaussian beam that has a waist of  $w_0\sim$~\SI{80}{\micro\meter} in radius ($1/\mathrm{e}^{2})$.  The sub-millimeter transverse dimensions of the cell allow for fast motional averaging \cite{Borregaard2016}, ensuring an integrated interaction between all atoms and the light. The microcell is positioned in a magnetic shield which contains coils producing a homogeneous magnetic bias field $B$ in the $x$-direction. The strength of the bias field splits the magnetically sensitive Zeeman levels by $|\omega_\text{S}|$, i.e., the Larmor frequency. Here, the Larmor frequency is in the range $\omega_\text{S}/2\pi \sim$ \SIrange{0.3}{1.5}{\mega\hertz}. The intrinsic (in the dark) spin damping rate at \SI{59}{\degree C} and \SI{1.5}{\MHz} Larmor frequency is $\gamma_{S0,\text{dark}}/2\pi=\SI{450}{\hertz}$. Circularly polarized resonant optical pumping lasers, co-aligned with the magnetic field, spin polarizes the ensemble \cite{Julsgaard2003MORS} to $\sim 80\%$. 

The remaining part of the setup in Figure~\ref{fig:setup} sets the phase sensitive control of the excitation and detection of the optical signal. We effectively generate an arbitrary polarization of the input light by combining two laser beams, here named local oscillator (LO) and modulation, with orthogonal polarizations on a polarization beam splitter (PBS) with a phase delay $\theta$. The modulation beam is phase modulated at a RF frequency $\omega_{\textrm{RF}}$ in a fiber electro-optic modulator (EOM) according to  $E_\text{EOM}\mathrm{e}^{i\theta}\mathrm{e}^{iG\sin\omega_\text{RF}t}$, for $G$ as the modulation strength. An arbitrary optical polarization state can be set by choosing a given $\theta$ and relative intensities of the beams. One of the output ports of the PBS is sent to the spin ensemble and the other is used for a polarization detection setup (phase lock).

We now describe the input polarization to the spin ensemble. The electric field of the laser light after the PBS is $E_\text{LO}(t)\hat{\mathrm{e}}_x + E_\text{EOM}(t)\hat{\mathrm{e}}_y \sim |E_\text{LO}|\hat{\mathrm{e}}_x + |E_\text{EOM}|\mathrm{e}^{i\theta}(1 + iG \sin \omega_{\textrm{RF}} t)\hat{\mathrm{e}}_y$, to the first order in $G$. Assuming $ |E_\text{EOM}|\ll|E_\text{LO}| $, the equivalent input Stokes parameters \cite{agarwal2003scheme} can be written, to the first order in $|E_\text{EOM}|$, as 
\begin{align}\label{eq:polarizationmodulation}
\begin{pmatrix}
    S_x^\text{in} \\
    S_y^\text{in} \\
    S_z^\text{in}
\end{pmatrix} =
\begin{pmatrix}
    |E_\text{LO}|^2/2 \\
    |E_\text{LO}||E_\text{EOM}| (\sin\theta+G\sin \omega_{\textrm{RF}}  t \cos\theta) \\
    |E_\text{LO}||E_\text{EOM}| (\cos\theta-G\sin \omega_{\textrm{RF}}  t \sin\theta)
\end{pmatrix}.
\end{align}
The phase $ \theta $ controls the relative contributions of circular and diagonal components, represented by $S_y^\text{in}$ and $S_z^\text{in}$, in the input polarization state. The DC (static) components will lead to a small rotation in the local oscillator's polarization; the AC components will induce CIFAR signal. In the linearized quadrature language, the effective AC input drive is written as $(-X_\textrm{L}^{\textrm{in}} \sin\theta+ P_\textrm{L}^{\textrm{in}}\cos\theta)G\sin \omega_{\textrm{RF}}  t$.

In the port that is directed to the phase lock output, the DC frequency interferometric signal, given in equation \eqref{eq:polarizationmodulation} is used for stabilizing $\theta$, the path length difference between the LO and the EOM arms, and therefore the input modulation to the spin ensemble. Deviations from the ideal phase shift induced by birrefringent elements, e.g., half wave plates in Figure~\ref{fig:setup}, leads to a further mixing between $S_y^\text{in}$ and $S_z^\text{in}$ in equation \eqref{eq:polarizationmodulation}, complicating the calibration of $\theta$. The same argument applies to setting the detection angle $\phi$. Experimentally, we first drive the spin oscillator with a RF magnetic field \cite{Julsgaard2003MORS} to set $\phi$. The detection half-wave plate is set to give the maximum photo-detected response to the applied magnetic field, which happens at $\phi=\SI{0}{\degree}$. Subsequently, we switch to a polarization modulation drive and find the effective $\theta=\SI{90}{\degree}$ when the spin response has a Lorentzian line-shape.

The polarization modulation at frequency $\omega_\text{RF}$ leads to a phase coherent interaction between the oscillator and light according to equation \eqref{eq:inout}. The signal is recorded by balanced polarimetry photodetection and processed by a lock-in amplifier phase-referenced to the drive, allowing us to extract the slowly varying amplitude $R$ and phase components. In a experimental protocol very similar to the one used in continuous wave Magneto-Optical Resonance Signal measurements \cite{Julsgaard2003MORS}, scanning $ \omega_\text{RF} $ around the resonant frequency $\omega_\text{S}$ at a rate much smaller than the spin damping rates, we ensure the steady-state performance and extract the signal of interest. To extract the useful parameters from the data, we implemented non-linear optimization and curve fitting routine to a two spin-modes model based in equation \eqref{eq:XLoutwithbb}.

For experimental implementations that operate in the pulsed regime \cite{julsgaard2001experimental,Wasilewski2010,VasilakisPRL2011}, in which the probing follows a spin-state preparation stage, the CIFAR signal can be extracted in a similar manner to the one prescribed in the continuous readout. For example, during a single repetition, a fixed-frequency polarization modulated probe can map out the time-evolving signal, thus obtaining a single point in the interference signal. Repeating the experiment with different drive frequencies allows for mapping a signal similar to those presented in the Results section. The data analysis therefore borrows the analysis discussed in the Theory section.

\section{Results}

\begin{figure}
    \centering
    \includegraphics[]{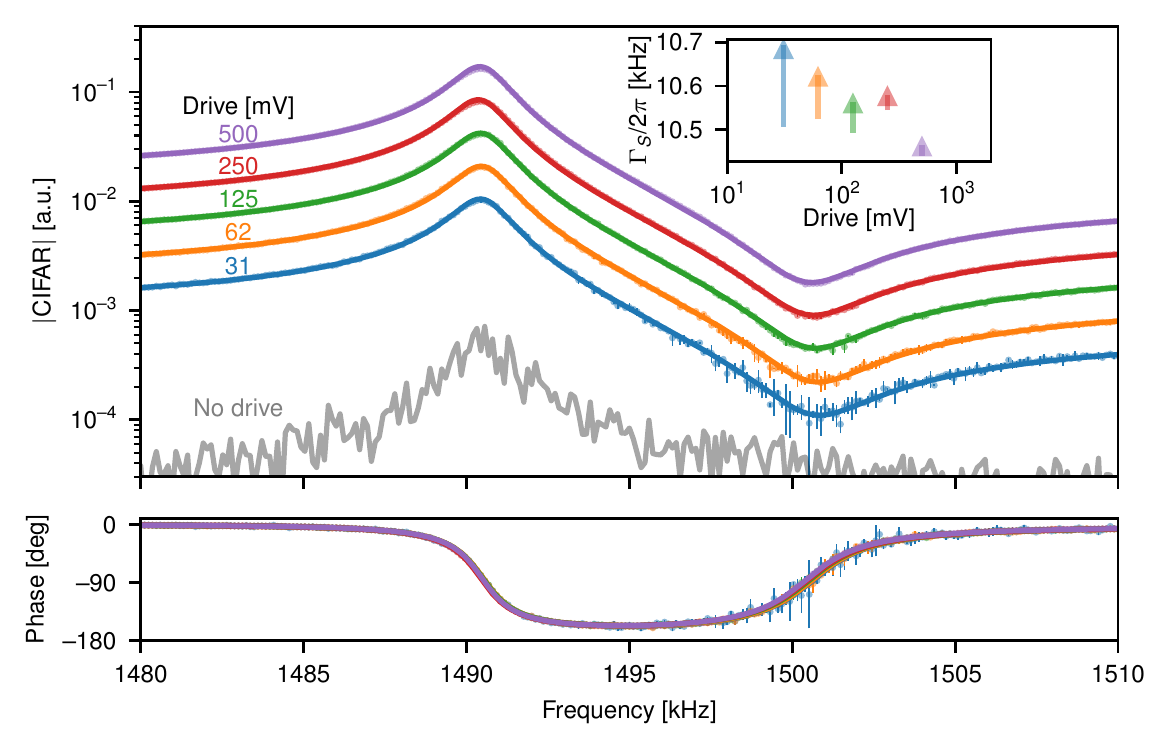}
    \caption{\textbf{CIFAR as a function of modulation amplitude.} CIFAR response amplitude (top) and phase (bottom) for different electrical EOM drive voltage $G$. The average of 3 scans (dots) is presented along with their statistical $1\sigma$ uncertainty error bars (vertical bars). The solid lines are the model fits to the individual curves. The grey line in the top panel is the measured response without any modulation at the input.  Inset: fitted readout rate and error bars as function of the drive voltage. For a discussion of the error bars, see the main text. }
    \label{fig:vsRFpower}
\end{figure}

We will now present experimental support to the model described in Section \ref{sec:model}. We fit the CIFAR model given in equation \eqref{eq:CIFARfinal} to the recorded data, present its performance on different experimental conditions and discuss the overall validity of the model. 

We start by the studying the response of spin oscillator to increasing modulation amplitudes $G$. In the data present in figures  \ref{fig:vsRFpower} to \ref{fig:Tseries}, we fix $\theta= \SI{45}{\degree}$, the probe power at $\SI{500}{\micro\watt}$, and cancel the non-linear quadratic Zeeman shift with tensor Stark shifts \cite{KasperPRA2009} by setting $\alpha\sim\SI{60}{\degree}$. In Figure \ref{fig:vsRFpower}, for each modulation amplitude, we record 3 scans and show the average as points, and the statistical $1\sigma$ uncertainty error bars (vertical bars). We double the amplitude starting from $\SI{31}{\mV}$ (in blue) until $\SI{500}{\mV}$ (in purple), showing $|\text{CIFAR}|$ (top panel) and the phase response (bottom panel), the amplitude and phase of the CIFAR signal, respectively. The grey line is the response of the spin oscillator to a shot noise-limited drive, in which the coherent polarization modulation is turned off. We see that the amplitude of the CIFAR signal follows the drive increase, doubling as the amplitude doubles. As the drive amplitude increases, the coherent response dominates the signal and the spread around the mean values decreases. 

\begin{figure}
    \centering
    \includegraphics[]{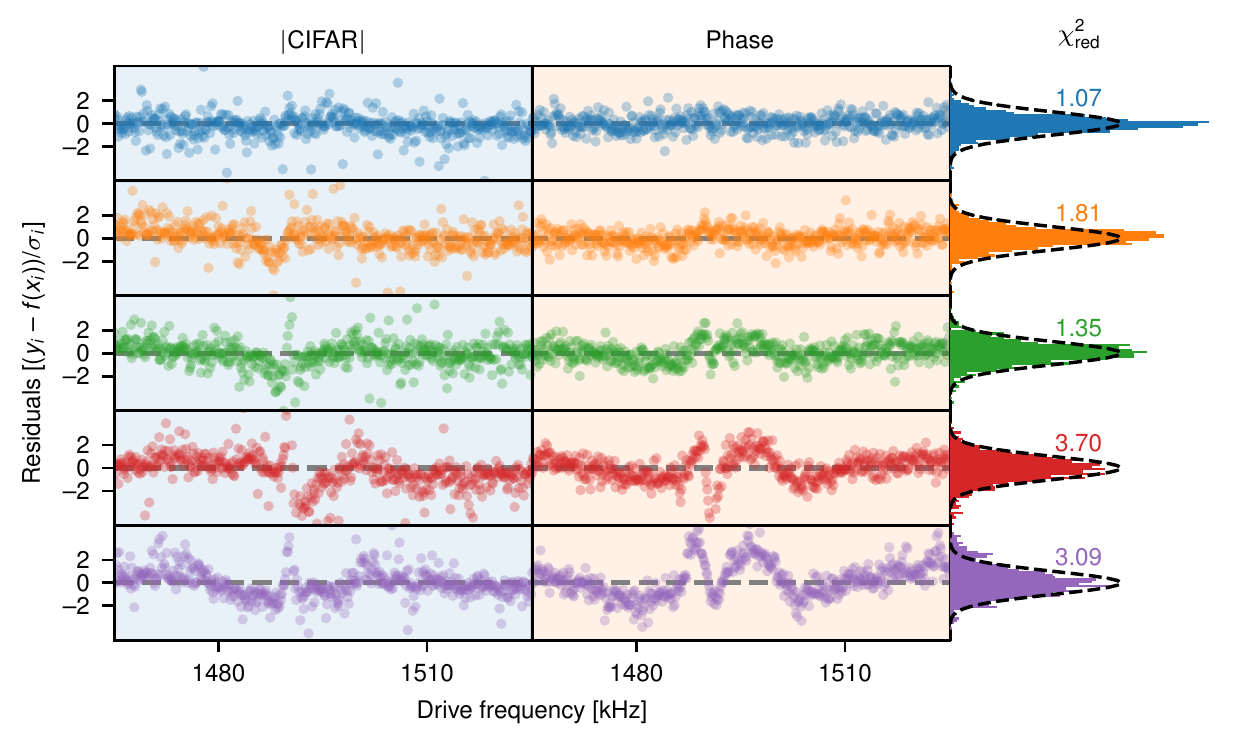}
    \caption{\textbf{Scaled fit residuals from Figure~\ref{fig:vsRFpower}.} Residuals between the model and data, both in CIFAR amplitude (left column) and phase (right column), for the various drive voltages shown in Figure~\ref{fig:vsRFpower}. In the right-most column we show the histogram of the residuals along with a unity width Gaussian curve (dashed lines) to guide the eye. We also print the reduced $\chi^2$. Some outliers are not shown.}
    \label{fig:residuals}
\end{figure}

The average traces for each drive amplitude in Figure \ref{fig:vsRFpower} is fit by the CIFAR model. The fits are displayed as solid lines, showing the good agreement to the measured amplitude and phase data. In the Figure \ref{fig:vsRFpower} inset, we show the readout rate $\Gamma_\text{S}/2\pi$ returned by the fitting routine as a function of the drive amplitude. For the $\SI{31}{\mV}$ drive amplitude, the value for the readout rate is $\Gamma_\text{S}/2\pi = \SI[parse-numbers=false]{10.685_{-0.18}^{+0.008}}{\kHz}$. For increasing drive, nonetheless, the trend is that the fitting routine returns smaller values. 

The asymmetric parameter errors are obtained with the \verb+conf_invertal+ function of the \verb+lmfit+ Python package. The function returns the parameter values for which $\chi^2=\chi_\mathrm{min}^2+1$, i.e., the interval containing the usual $68.27 \%$ probability, which for a Gaussian parameter error corresponds to the $1\sigma$ uncertainty. Similar results was obtained  by Markov Chain Monte Carlo optimization (not shown). The asymmetry arises due to a strong correlations between $\Gamma_\mathrm{S}$ and other fit parameters, mainly the parameter describing the overall response $\propto G$.

We further note that successfully fitting the model to the data relies on reliably ascribing errors to the individual data points; the individual traces spans 2 orders of magnitude, and failing to account for this in the optimization routine leads to discrepancies in either the peak or valley of the traces. Curiously, the data errors largely inherit the shape of the undriven atomic ensemble, i.e., a Lorentzian centered on the spin frequency (not shown). This places the condition that to obtain good fit values, all measurements must be repeated a number of times, to obtain trustworthy statistics.

Looking at the fitting residuals for the different traces, presented in Figure \ref{fig:residuals}, we see that structured deviations between model and data appear as the spin oscillator is driven with larger amplitudes. The residuals to the traces with drive voltages above $\SI{250}{\mV}$, shown in red and purple points, present significant deviations from our model. We believe that the deviations from our model appear as we start to drive the spin system significantly away from the small oscillation amplitude limited approximation that leaves the system beyond the linearized oscillator model. We have, therefore, experimentally found the limit on the drive strength that our model can describe and the driving amplitude regime that must be used to return results free of systematic effects.

In Figure \ref{fig:Tseries} we present the dependence of the readout rate on $J_x$, the mean spin length. According to equation \eqref{eq:readoutrate}, $\Gamma_S\propto J_x$. The spin length is controlled by the temperature of the vapor cell, which sets number of atoms. When heated, the cesium vapor pressure increases \cite{steck2003cesium}. For the data on Figure \ref{fig:Tseries}, we record CIFAR scans while the cell is heated from \SI{\sim 34}{\degree C} (blue points) to \SI{\sim 59}{\degree C} (purple points). The solid lines are fits to the data, with the frequency axis shifted according to $\Delta_\text{RF}$ and re-scaled to the returned spin linewidth $\gamma_\text{S}$. The extracted readout rate increases from $\SI{1.1}{\kHz}$ to $\SI{10.0}{\kHz}$. As the temperature, and consequently $J_x$, is increased, the peak signal increases and the minimum is shifted up in frequency. Importantly, the frequency detuning $\Delta_\text{RF}$ for which the CIFAR signal is minimal follows the readout rate $\Gamma_\text{S}$, as shown in the inset. There is, approximately, a one-to-one correspondence between $\Gamma_\text{S}/\gamma_\text{S}$ and the frequency of the CIFAR signal minimum value, as shown by the line with slope 1 (solid line). Therefore, by choosing the input modulation type $\theta=\SI{45}{\degree}$,  an approximate readout rate can be easily extracted from the CIFAR signal as the frequency difference between the maximum and minimum of the trace. We also note that at the highest temperature setting ($\sim$\SI{59}{\degree C}), with linewidth $\gamma_{\text{S}0}/2\pi=\SI{1.3}{\kHz}$ 
and estimated spin thermal occupation $n_\text{S}\sim 0.75$ \cite{rodrigosthesis}, we have $\Gamma_\text{S}/\gamma_\text{S}\sim 7$ and estimate $C_\text{q}\sim 3$, indicating that the spin oscillator is strongly coupled to light.

\begin{figure}
    \centering
        \includegraphics[]{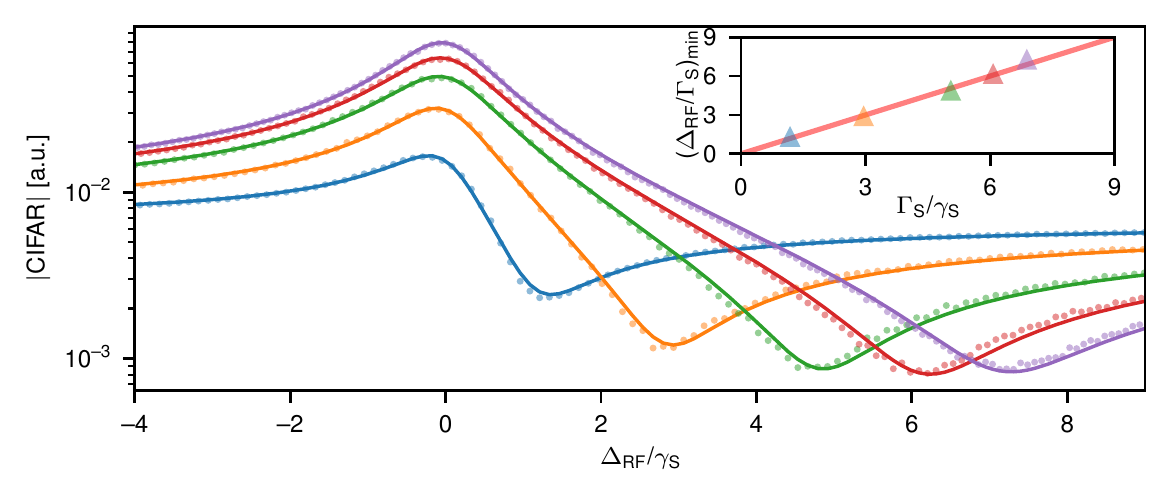}
    \caption{\textbf{CIFAR scans for different $\mathbf{\Gamma_S}\pmb{/}\pmb{\gamma_S}$.} We vary the readout rate by changing the temperature of the cell from \SI{\sim 34}{\degree C} to \SI{\sim 59}{\degree C}. Inset: The location of the minimum of the CIFAR response in units of $\Delta_\text{RF}/\gamma_\text{S}$ as a function of the normalized readout rate $\Gamma_\text{S}/\gamma_\text{S}$. Solid line: line with slope 1.}
    \label{fig:Tseries}
\end{figure}

In Figure \ref{fig:vsZeta}, we present the CIFAR signal for different strengths of the tensor coupling parameter $\zeta_\text{S}$. For a given detuning from the atomic resonance $\Delta$, it is modified by selecting $\alpha$, the angle between the LO linear polarization and the magnetic field $B$ direction. According to equation \eqref{eq:tornado}, the angle $\alpha=\SI{45}{\degree}$ turns off the tensor coupling. For this experiment, we reduced the spin resonance frequency to $\omega_\text{S}\sim\SI{400}{\kHz}$ to avoid non-linear Zeeman splitting \cite{Julsgaard2003MORS}, the probe power was set to $\SI{250}{\micro\watt}$ to reduce probe-induced power broadening, and the temperature to $T=\SI{55}{\degree}$. In Figure \ref{fig:vsZeta}, we show the amplitude of the CIFAR signal for $\theta=\SI{45}{\degree}$ (top panel) and $\theta=\SI{90}{\degree}$ (bottom panel).  The data for $\alpha=\{\SI{0}{\degree},\SI{45}{\degree},\SI{90}{\degree}\}$ are shown in blue, orange and green dots, respectively. The choice of $\theta=\SI{45}{\degree}$ gives responses similar to those presented in Figure \ref{fig:vsRFpower}. For this data set, we have $\Gamma_\text{S}=\SI{4.9}{\kHz}$. The setting $\theta=\SI{90}{\degree}$, nonetheless, gives a rather different picture. According to equation \eqref{eq:CIFARfinal}, the detected signal goes as
\begin{align}
     |\text{CIFAR}(\theta=\SI{90}{\degree},\phi=0)/G|^2 &= \left|1-2 \Gamma _\text{S} \zeta _\text{S} \left(\tfrac{\gamma _\text{S}}{2}-i \omega_\text{RF} \right)\chi_\text{S}(\omega_\text{RF})\right| ^2 \nonumber \\
     &\sim 1-  \dfrac{\zeta _\text{S} \Gamma _\text{S}  \gamma_\text{S}}{\Delta_\text{RF}^2 + (\gamma_\text{S}/2)^2},
\end{align}
where in the last passage we used the high-Q ($\gamma_\text{S}\ll \omega_\text{S}$ and $\omega_\text{RF}\sim \omega_\text{S}$), and the small tensor coupling ($\zeta_\text{S}\ll 1$) limit. For this configuration, the CIFAR is dominated by the constant term, 
since we mostly detect the input modulation. 
Near the spin resonance, the oscillator will add ($\zeta_\text{S}<0$) or remove ($\zeta_\text{S}>0$) signal according to the tensor coupling sign. The obtained tensor coupling parameters are $\zeta_\text{S}=-0.045\pm0.002$ and  $\zeta_\text{S}=0.040\pm0.003$ for $\alpha=\SI{0}{\degree}$ and $\alpha=\SI{90}{\degree}$, respectively. For reference, the expected tensor parameter for a perfectly spin polarized ensemble is $|\zeta_\text{S}^\text{th}|=0.053$. For $\alpha=\SI{45}{\degree}$ the spin contribution is, according to our theory, null; the returned value is $\zeta_\text{S}=0.000\pm0.001$.

\begin{figure}
    \centering
        \includegraphics[]{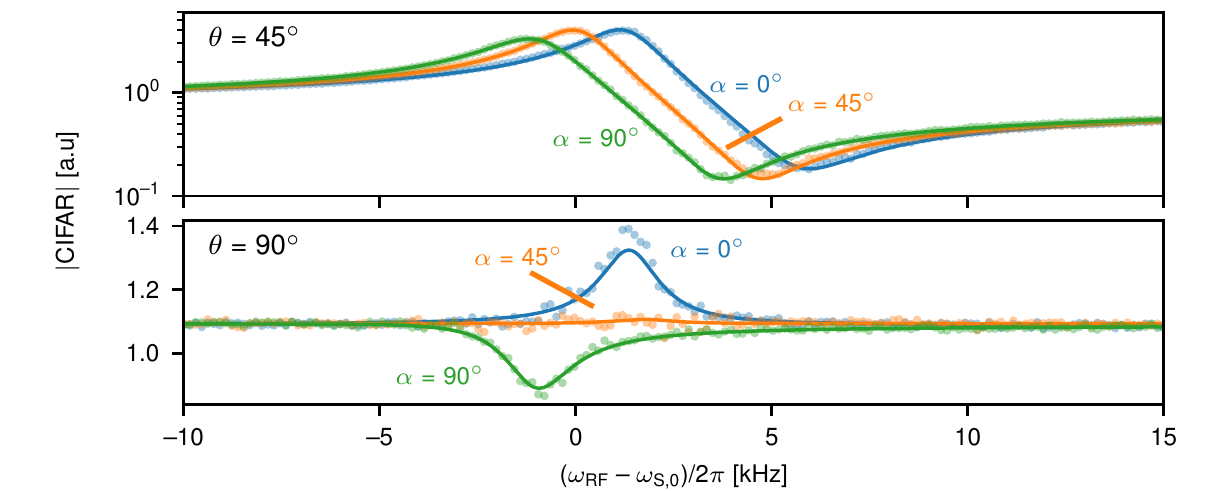}
    \caption{\textbf{CIFAR signal for different tensor coupling parameters $\zeta_\text{S}$.} The overall response of the spin oscillator to light depends on $\zeta_\text{S}$, here controlled by the angle $\alpha$ between the input linear polarization LO and the direction of the magnetic bias field $B$. The CIFAR signals for input modulation with $\theta=\SI{45}{\degree}$ (top panel, logarithmic scale) and with $\theta=\SI{90}{\degree}$ (bottom panel, linear scale), and $\alpha=\{\SI{0}{\degree},\SI{45}{\degree},\SI{90}{\degree}\}$ are shown in blue, orange and green, respectively.} 
    \label{fig:vsZeta}
\end{figure}

In our last study we present the broadband spin contributions to the CIFAR signal. The measurements presented in Figure \ref{fig:widecifar} are taken in the same experimental conditions as the data in Figure \ref{fig:vsRFpower}, but now scanning the drive tone in a $\sim\SI{600}{\kHz}$ band around $\omega_\text{S}$. The CIFAR signal amplitude (top panels) and phase (bottom panels), including the model fits, are shown for $\theta \in \{\SI{-45}{\degree}, \SI{0}{\degree}, \SI{45}{\degree}\}$ in blue, orange and green, respectively. Apart from the symmetric changes in the response as $\theta$ is changed from $\SI{-45}{\degree}$ to $ \SI{45}{\degree}$, the $\theta \sim  \SI{0}{\degree}$ amplitude and phase responses display characteristic features of a broadband spin response. The broadband spin response can be clearly seen by setting $\Gamma_\text{S}=0$ (dashed orange line) or in our full model fit (dark orange line). The light orange line corresponds to the predicted response of the spin oscillator in the case $\Gamma_\text{S,BB}=0$. The broadband spin response, having a damping rate $\gamma_\text{S,BB}/2\pi=\SI{0.93}{\MHz}$, couples to the drive with rate $\Gamma_\text{S,BB}/2\pi=\SI{33.4}{\kHz}$, distorting the phase response  and adding a pedestal to the detected amplitude. Remarkably, although having a complex origin \cite{Shaham2020}, the broadband response is qualitatively well described by a single effective mode.

\begin{figure}
    \centering
        \includegraphics[]{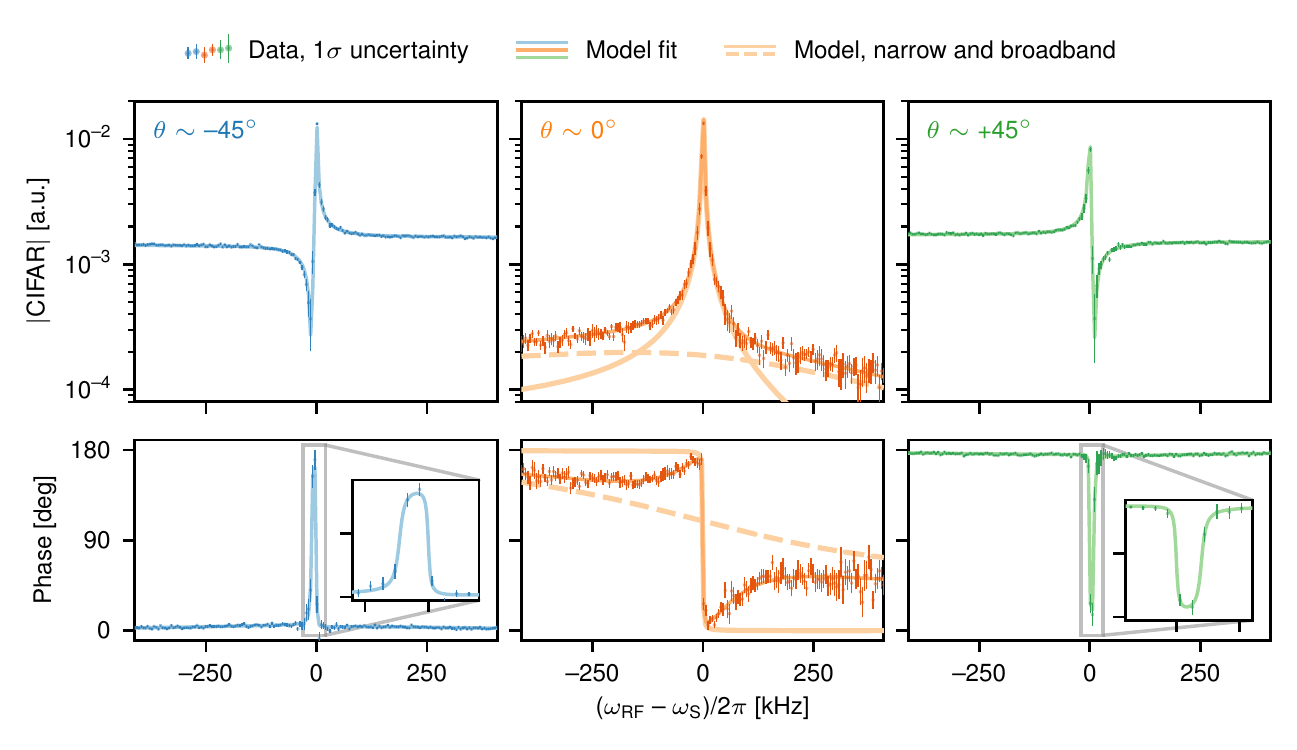}
    \caption{\textbf{Coherent interference between the responses of the narrow and broadband spin modes}. CIFAR response amplitude (top row) and phase (bottom row) as a function frequency detuning for three different modulation phases, $\theta \in \{\SI{-45}{\degree}, \SI{0}{\degree}, \SI{45}{\degree}\}$. The data was taken under the same experimental conditions as the $\SI{62}{\mV}$ drive trace (orange curve) in Figure~\ref{fig:vsRFpower}. For $\theta=\SI{0}{\degree}$ we plot the fit result evaluated with the broadband readout rate set to $\Gamma_\mathrm{S,BB}=0$ (solid light orange curves, top and bottom panels) and narrowband readout rate $\Gamma_\mathrm{S}=0$ (dashed light orange curves).}
    \label{fig:widecifar}
\end{figure}

\ifoptionfinal{}{
\section{Conclusion}
}{}

In summary, we have presented a novel approach for calibrating the light-matter interaction between off-resonant optical beams and collective spin systems, the CIFAR technique. Ex\-peri\-mentally, the calibration method relies only on applying a known input modulation and detecting a known optical quadrature, the variables parametrized by $\theta$ and $\phi$, respectively. Supplied with the input-output relations of the spin-light interaction, a simple procedure for determining the interaction parameters, among those most importantly the readout rate $\Gamma_\text{S}$, is described. Fitting the recorded signal to the full model provides a full characterization of the system parameters. The technique does not rely on knowing the photo-detection efficiency or the ensemble spin polarization. We have verified the good agreement between data and the CIFAR method by continuously probing a strong coupled spin oscillator prepared in a warm cesium atomic vapour.

Theoretical refinement of our model can be envisioned with the consideration of the full Zeeman structure of the ground state manifold \cite{colangelo2013quantum} ($F=4$, in the present example). With that, the model will be able to account for the non-unity spin polarization and return consistent values for the tensor parameter. Furthermore, in the case of large excursions by the transverse angular momentum variables induced by the drive light, going beyond the linearized regime can account for the mismatch between theory and data shown in our residuals analysis.

The technique here presented is also a powerful method for studying the coupling of light to the spin modes under diffusion and spatial averaging \cite{Shaham2020}. Our works provides an evidence for the coherent coupling and classical back-action of the short-lived spin modes with light, as opposed to previous observations of just broadband spin noise \cite{Tang2020}. The CIFAR technique also paves the way for probing and engineering the optical coupling of higher order spin modes to light, a source of inefficiencies and unwanted noise in quantum limited measurements. By preparing the optical field in a suitable spatial mode, the multimode capabilities of the spin-light platform can be utilized.

\section*{Funding}
Villum Fonden (Villum Investigator Grant 25880); European Research Council (QUANTUM-N); John Templeton Foundation.

\section*{Acknowledgments}
The authors acknowledge Jörg H. Müller and Emil Zeuthen for enlightening conversations. M.P. is also supported by the MAB/2018/4 project "Quantum Optical Technologies", carried out within the International Research Agendas program of the Foundation for Polish Science co-financed by the European Union under the European Regional Development Fund.

\bibliography{references.bib}

\end{document}